\documentclass[conference]{IEEEtran}

\usepackage{cite}

\ifCLASSINFOpdf
  \usepackage[pdftex]{graphicx}
\else
\fi

\usepackage{amsmath,amssymb,amsfonts}
\usepackage{algorithmic}
\usepackage{textcomp}
\usepackage{xcolor}
\usepackage{hyperref}
\def\BibTeX{{\rm B\kern-.05em{\sc i\kern-.025em b}\kern-.08em
    T\kern-.1667em\lower.7ex\hbox{E}\kern-.125emX}}

\usepackage{csquotes}
\usepackage{multirow}
\usepackage{enumitem}
\usepackage{xurl}
\usepackage[online]{threeparttable}
\usepackage{pifont}
\newcommand{\cmark}{\ding{51}}%
\newcommand{\xmark}{\ding{67}}%
\newcommand{\amark}{\ding{107}}%
\newcommand{\upmark}{\ding{115}}%
\newcommand{\lnmark}{\ding{116}}%
\newcommand{\smark}{\ding{73}}%

\begin{document}

\title{A First Look at Digital Rights Management Systems for Secure Mobile Content Delivery}

\author{\IEEEauthorblockN{Amir Rafi}
\IEEEauthorblockA{\textit{Information Security Group} \\
\textit{Royal Holloway, University of London}\\
Egham, Surrey, United Kingdom\\
amir.rafi.2019@live.rhul.ac.uk}
\and
\IEEEauthorblockN{Carlton Shepherd}%
\IEEEauthorblockA{\textit{School of Computing}\\
\textit{Newcastle University}\\
Newcastle-upon-Tyne, United Kingdom\\
carlton.shepherd@ncl.ac.uk}
\and
\IEEEauthorblockN{Konstantinos Markantonakis}
\IEEEauthorblockA{\textit{Information Security Group} \\
\textit{Royal Holloway, University of London}\\
Egham, Surrey, United Kingdom\\
k.markantonakis@rhul.ac.uk}
}

\maketitle

\begin{abstract}
Digital rights management (DRM) solutions aim to prevent the copying or distribution of copyrighted material. On mobile devices, a variety of DRM technologies have become widely deployed. However, a detailed security study comparing their internal workings, and their strengths and weaknesses, remains missing in the existing literature. In this paper, we present the first detailed security analysis of mobile DRM systems, addressing the modern paradigm of cloud-based content delivery followed by major platforms, such as Netflix, Disney+, and Amazon Prime. We extensively analyse the security of three widely used DRM solutions---Google Widevine, Apple FairPlay, and Microsoft PlayReady---deployed on billions of devices worldwide. We then consolidate their features and capabilities, deriving common features and security properties for their evaluation. Furthermore, we identify some design-level shortcomings that render them vulnerable to emerging attacks within the state of the art, including micro-architectural side-channel vulnerabilities and an absence of post-quantum security. Lastly, we propose mitigations and suggest future directions of research.
\end{abstract}


\section{Introduction} \label{intro}
Video streaming consumes over half of the global Internet bandwidth \cite{noauthor_global_2022}. The market share of over-the-top (OTT) media services---those offered directly to viewers via the Internet, bypassing cable, broadcast, and satellite television---is projected to reach USD 139 billion in 2028 \cite{noauthor_over_2021}. Current trends in OTT media broadcasting indicate the migration of all workflows to the cloud, from production and packaging to delivery \cite{fachot_cyber_2019-1}.  The OTT model, however, poses some significant challenges from a security standpoint. Media content is received and displayed by devices that are under the full control of customers, i.e.\ mobile phones and tablets.  Cloud-based content providers, such as Netflix and Disney+, rely on DRM to prevent copyright infringements and unauthorised consumption of content. A DRM system restricts content usage by delivering encrypted content to authorised devices and controlling its decryption; legitimate users are thus potential adversaries in a typical DRM threat model. 

The current DRM landscape consists of proprietary solutions causing a fragmentation in the industry; content providers are thereby compelled to implement multiple DRM systems to support multiple platforms. Today's mobile platforms are shipped with proprietary DRM systems, which are integrated into the application or operating system itself, such as Android (Widevine), iOS (FairPlay), and Microsoft Windows (PlayReady). Android and iOS have a global OS market share of 72.11\% and 27.22\% respectively for mobiles; 47.54\% and 52.38\% respectively for tablets; and Windows OS is used on 76.31\% of desktops \cite{noauthor_operating_2022}. Widevine, for instance, is available on 5 billion devices worldwide~\cite{noauthor_widevine_nodate-2}.  

\subsection{Contributions}

Since 1998, the Digital Millennium Copyright Act (DMCA) has hindered security researchers from analysing DRM systems. Recent literature describes a number of exploits, reverse engineering of code obfuscation techniques used for protecting low quality content, and analysis of OTT apps that rely on DRM protection. As a result, despite their widespread use, the understanding of various DRM technologies remains highly fragmented. This paper aims to bridge this gap. We systematically evaluate the security of mainstream DRM systems using an approach and criteria that analyses such systems holistically. Such an evaluation remains forthcoming in the existing literature. Our contributions can thus be summarised as follows:

\begin{itemize}
\item{To the best of our knowledge, we present the first domain-specific security evaluation of mobile-based DRM systems, considering the  security of both the content and user devices.}



\item{Based on a comprehensive analysis of public domain material, we identify probable attack vectors and present a comparative security analysis (\S\ref{analysis}) of Widevine, FairPlay and PlayReady DRM systems.} 

\item{We discuss mitigations (\S\ref{mitigations}) for security vulnerabilities discovered during our analysis and suggest future directions of research (\S\ref{future}).}
\end{itemize}

\subsection{Scope} \label{scope}
A DRM system can be deployed on user platforms as a separate application, as a part of the web browser, or as a part of the OS, with optional hardware protection. From a deployment perspective, DRM solutions may be found in the form of an obfuscated library (see~Widevine L3~\cite{WidevineDRMArchitecture2017a}) or, increasingly, within a trusted execution environment (TEE) that prevents kernel-mode adversaries from accessing relevant assets, such as content decryption keys. Broadly speaking, the most sophisticated DRM implementations are integrated into the OS, leveraging kernel-level security mechanisms, and hardware-assisted protections offered by TEEs.

The scope of this security analysis is limited to DRM deployments implemented as a part of the OS and using TEEs for hardware-level protection. In this work, we focus on DRM systems implemented on mobile platforms, comprising smartphones, tablets, and laptop devices possessed by viewers of media content (i.e.\ the general public). DRM solutions are notoriously closed-source in nature, which restricts a complete understanding of their characteristics. To conduct this study, we thus rely on publicly available materials pertaining to Widevine, FairPlay and PlayReady to analyse their security. We take a security-oriented approach to this work; issues relating to privacy and content watermarking are demeed outside the scope of this work.




\section{Background \& Related Work}  \label{background} 

The DMCA was signed in to the United States law in 1998 to discourage content piracy, and amended as Title 17 \S1201~\cite{noauthor_17_1999} prohibiting the circumvention measures for protecting copyrighted material. Although piracy continued to be conducted, the DMCA restricted security researchers from analysing DRM systems~\cite{noauthor_digital_nodate-4}. 
In the 2000s, DMCA takedowns were used against security researchers who identified issues with the High Bandwidth Digital Content Protection (HDCP) scheme developed by the Intel Corporation~\cite{ferguson_censorship_2001,crosby_cryptanalysis_2002}. By 2016, the DMCA anti-circumvention claims had been raised in nearly 50 reported cases in the United States federal courts~\cite{noauthor_list_2016}. The exemptions to DMCA were expanded by the Copyright office and the Library of Congress in 2018, allowing security researchers greater freedom to analyse DRM systems.

In recent years ($>$2012), the growth of OTT mobile streaming has led to new generation of DRM systems, prompting researchers to analyse their claimed security assurances. The MITRE CVE database lists 53 CVE records covering different security issues in Widevine and PlayReady since 2014.\footnote{There are no CVE records for FairPlay, although work has identified some significant security weaknesses~\cite{plafke_chinese_2013, chau_why_2018-1}.} In 2013, Wang et al.~\cite{wang_steal_2013} introduced an automated technique to remove DRM protection from content protected by PlayReady. This technique tracks the flow of content between memory buffers in real-time during playback, identifies where the content is decrypted, and copies the decrypted content, which is then reconstructed into a new media file without any DRM protection. The authors demonstrated its effectiveness against Amazon Instant Video, Hulu, Netflix, and Spotify.

D'Orazio and Choo \cite{dorazio_adversary_2016} proposed a DRM adversary model in 2016 to formalise potential attackers' capabilities and to evaluate DRM protection on iOS devices. They highlighted vulnerabilities resulting from a lack of public-key pinning and server-side verification. Since 2013, a technique known as \enquote{FairPlay MITM (Man-In-The-Middle)} \cite{plafke_chinese_2013} exploited vulnerabilities in FairPlay to propagate pirate iOS apps. In 2016, Xiao \cite{xiao_acedeceiver_2016-1} reported how FairPlay MITM could be used to spread malware on iOS devices.

Against Widevine, a vulnerability \cite{chirgwin_googles_2016-1} was discovered in 2016 that allowed access to a local cache containing decrypted version of protected content. This vulnerability was tested against Amazon Prime and Netflix, although other DRM systems may also be vulnerable. In 2018, Chau et al. \cite{chau_why_2018-1} analysed 141 content delivery apps on the Android platform and discovered vulnerabilities that not only compromised the DRM-protected content but also put users' security and privacy at risk. The vulnerabilities were caused by insecure policy delivery and bootstrapping, client-side policy enforcement, and reusing encryption keys; the whole Amazon Music collection of 40 million songs was found to be encrypted with the same key. The work evaluated the apps that \emph{use} DRM systems for content protection, but not the DRM systems themselves.

In 2019, Buchanan \cite{buchanan_breaking_2019} used differential fault analysis to recover keys for protected content using Widevine level 3 (L3)~\cite{noauthor_googles_2019-1} (see \S\ref{widevine}). Most GitHub repositories containing this proof of concept code were removed by Google's DMCA take-down \cite{noauthor_dmca_2020} in 2020. More recently, Zhao~\cite{zhao_wideshears_2021-1} broke in to the Widevine L1 security level on the Qualcomm TEE, an Android TrustZone implementation. The L1 is the most secure implementation of Widevine, which uses TEEs on user devices to protect content and encryption keys, and to securely decrypt content. Zhao recovered the Widevine keybox from the trusted storage of QTEE by breaking the ASLR (Address Space Layout Randomization). 


Based on an analysis of OTT apps including Netflix, Hulu and Showtime, Patat et al.~\cite{patat_wideleak_2022} revealed that most apps use the same cryptographic keys for a particular media item across all users. Moreover, these apps support vulnerable legacy devices, which can be exploited to compromise the cryptographic keys. In 2022, they used a vulnerable legacy device to recover the Widevine L3 root of trust (CVE-2021-0639) by bypassing the obfuscation used to conceal cryptographic keys in software~\cite{patat_exploring_2022}. The authors exploited design decisions for supporting legacy devices; however, they only compromised low-quality content made available by Widevine L3. Patat et al.~\cite{patat_wideleak_2022} raised privacy concerns over the device identifiers in the Widevine protocol, which can be used for third-party user tracking. Interestingly, their responsible disclosure revealed that Netflix was unaware that it does not protect audio tracks.  The work investigated OTT apps and their compliance with Widevine guidelines on Android devices; it did not directly analyse the security of Widevine.



\section{Security Design of DRM Systems} \label{security goals}

\subsection{Security Architecture of Google Widevine} \label{widevine}
\begin{figure}
\centering
 	\includegraphics[width=\linewidth]{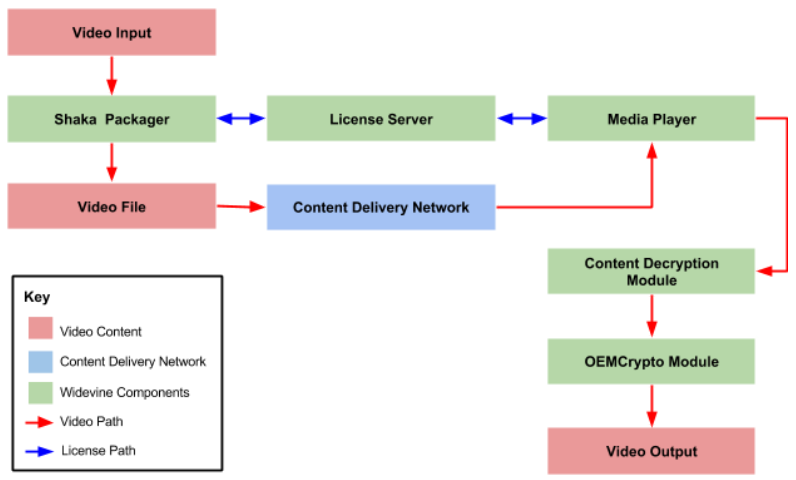}
 	\caption{Widevine DRM architecture \cite{WidevineDRMArchitecture2017a}.} 
 	\label{fig:widevinedesign} 
  \vspace{-0.3cm}
\end{figure}

\begin{figure}
\centering
 	\includegraphics[width=\linewidth]{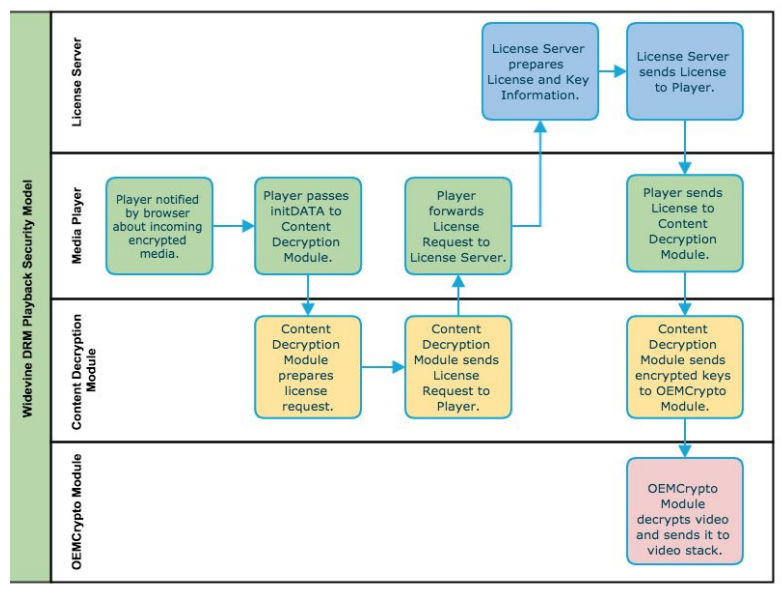}
 	\caption{Widevine playback security model \cite{WidevineDRMArchitecture2017a}.} 
 	\label{fig:widevinedecrypt} 
  \vspace{-0.3cm}
\end{figure}

Widevine is the primary DRM implementation for Android devices, which is deployed as a part of the operating system and it can also be deployed as a browser- or application-based implementation. It is used by various content providers, including Google Play, Amazon Prime, Netflix, YouTube, BBC, Hulu, Disney+ and Spotify \cite{GoogleWidevineDigital2021,kimWhatGoogleWidevine2019}. Its main components comprise the Shaka packager, the Widevine license server, a supported media player, the Content Decryption Module and the OEMCrypto module~\cite{WidevineDRMArchitecture2017a}. Figures \ref{fig:widevinedesign} and \ref{fig:widevinedecrypt} demonstrate how the components interact together to deliver, decrypt, and playback content on user devices. We describe each component as follows.

\subsubsection{Security Levels}

Widevine uses three different security levels: L1 to L3. L1 is deemed the most secure and uses the TEE for both cryptography and media processing. The L2 security level uses the TEE for cryptography but not media processing, returning decrypted content to video decoder in the non-secure world for display. The L3 security level is the least secure and uses only software-based protection on devices that do not have a TEE \cite{WidevineDRMArchitecture2017a}. All Widevine security levels require signed system images to be loaded on devices by a secure bootloader. The L1 and L2 security levels require the Widevine components implemented in the TEE to be included in the bootloader chain of trust; L1 also requires the components used for the secure video path to be included \cite{leeWidevineDRMDevices2018a}.  The L1 and L2 security levels also require the Widevine Keybox and the OEMCrypto API to be implemented on the device for key management and content decryption. The Keybox is encrypted with a unique factory-provisioned and device-specific key \cite{leeWidevineDRMDevices2018a}.

\subsubsection{Shaka Packager}
This is Widevine's open-source packager that uses the Dynamic Adaptive Streaming over HTTP (DASH) protocol \cite{WidevineDRMArchitecture2017a, stockhammerDynamicAdaptiveStreaming2010} for optimised streaming over a dynamic bandwidth; HTTP Live Streaming (HLS) \cite{HTTPLiveStreaminga} is also supported. The DASH protocol converts content into fragmented MP4 files containing multiple versions of segments with the same length but different resolutions and bitrates. A media presentation description (MPD), or a manifest file, describes the different resolutions and bitrates for the content. The media player can use the manifest to request segments of different resolutions depending on the available bandwidth.

\subsubsection{Widevine Encryption}
The Shaka packager encrypts the media segments during packaging. Widevine supports the CENC (Common Encryption) \cite{ISOIEC2300172016} and recommends the ISOBMFF (ISO Base Media File Format) standard; FairPlay Streaming encryption \cite{FairPlayStreamingHTTP} and the WEBM format are also supported. Widevine uses the Encrypted Media Extensions (EME)~\cite{dorwinEncryptedMediaExtensions2017} and Media Source Extensions (MSE)~\cite{wolenetzMediaSourceExtensions2022} to allow playback of encrypted media.

\subsubsection{License Server}
Widevine provides a cloud-based license server that uses a request-response protocol over HTTPS to issue content licenses to the Content Decryption Module.

\subsubsection{Content Decryption Module (CDM)}
This is a proprietary component of Widevine on user devices, which creates encrypted license requests that are relayed to the license server by the media player application. The CDM receives encrypted license responses from the license server, through the media player, and uses the OEMCrypto module to decrypt them.

\subsubsection{OEMCrypto Module}
Decrypts the license responses to extract the encryption keys, which are used to decrypt the encrypted media content within the TEE of the user's device. OEMCrypto is implemented by the system-on-chip vendor for Widevine L1, or by Widevine (Google, Inc.) for L3.

\subsubsection{Widevine Playback Security}
The Shaka packager produces encrypted media segments and a corresponding manifest. The encryption keys are sent to the license server and the encrypted media segments are streamed to the media player on user devices through the Content Delivery Network (CDN). The application receives a manifest and encrypted content from the CDN, extracts the initialisation data (InitData), and sends it to the media player. The media player forwards the InitData to the CDM to create a license request; the player sends the (encrypted) licence request produced by the CDM to the license server, which returns an (encrypted) license response back to the CDM. The license server authenticates the media player before responding to any requests. Next, the CDM sends the license response and media to the OEMCrypto module in the TEE to decrypt the content; OEMCrypto sends the decrypted content to the video stack for display. 


\subsection{Security Architecture of Apple FairPlay} \label{fairplay}
Apple FairPlay, implemented in Apple iOS and tvOS, is based on the FairPlay Streaming (FPS) specification \cite{FairPlayStreamingHTTP}. FPS provides secure delivery of encryption keys to devices and allows secure playback of encrypted content, which is provided using the HTTP Live Streaming (HLS) protocol \cite{HTTPLiveStreaminga}. We note that the Safari browser implementation on macOS uses FPS with an EME interface~\cite{FairPlayStreamingOverview2016a}.

\begin{figure}
\centering
 	\includegraphics[width=0.8\linewidth]{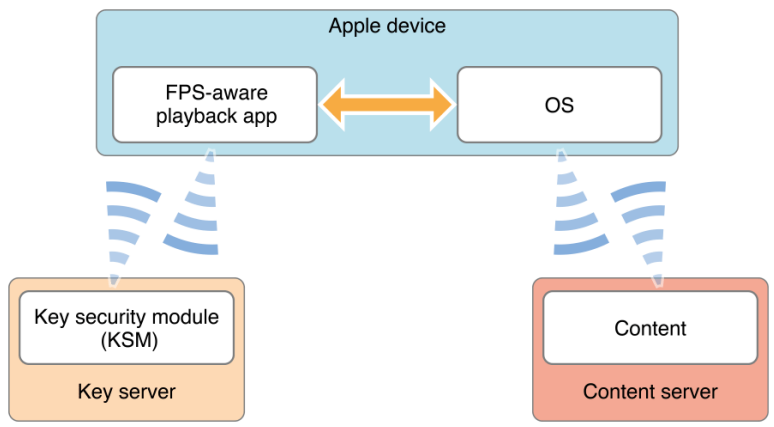}
 	\caption{Interactions between FPS and content provider’s servers \cite{FairPlayStreamingOverview2016a}.} 
 	\label{fig:fpscomponents} 
  \vspace{-0.3cm}
\end{figure}

FPS requires content providers to implement their own content server, key (license) server and a player application; these must meet FPS requirements to be compatible with the FPS framework \cite{FairPlayStreamingOverview2016a}.  The content providers also need a public key infrastructure and public key certificates for RSA keys that are used to encrypt the license requests \cite{FairPlayStreamingOverview2016a}.  FPS authenticates the key server but requires content providers to source their own mechanisms for the player application to be authenticated by the key server \cite{FairPlayStreamingOverview2016a}. FPS does not include content encryption and packaging mechanisms. Content providers are required to use the HLS streaming protocol for content delivery, and AES-128 CBC mode for encryption; video content is encrypted on a per frame basis, using a unique initialization vector (IV) for each encryption, and the audio content is fully encrypted on a per sample basis. Content providers handle the encryption keys and the IVs, as well as their associations with the content \cite{FairPlayStreamingOverview2016a}. Figure \ref{fig:fpscomponents} shows interactions between FPS on user devices and the content provider's key and content servers. FPS is integrated into the device OS and creates a server playback context (SPC) to request content encryption keys from the key server. The key server responds with the content key context (CKC). FPS extracts the encryption keys from the CKC and decrypts the content for display \cite{FairPlayStreamingOverview2016a}.

\subsubsection{FairPlay Playback Security}
\begin{figure}
\centering
 	\includegraphics[width=\linewidth]{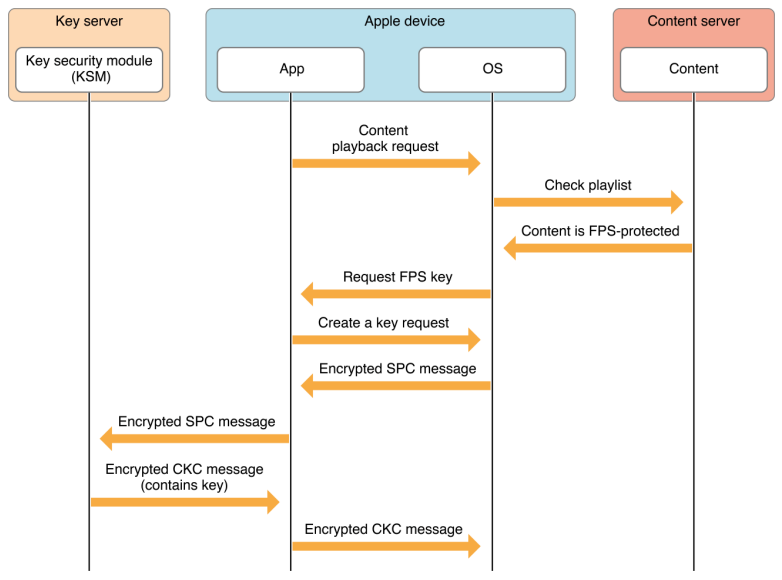}
 	\caption{Sequence of data flow in FPS~\cite{FairPlayStreamingOverview2016a}.} 
 	\label{fig:fpsprocesses} 
  \vspace{-0.3cm}
\end{figure}

Figure \ref{fig:fpsprocesses} shows the sequence of data flow during content playback. The player application sends a playback request to FPS on the device, which refers to the content manifest playlist on the content server and obtains the encrypted content. The FPS requests the player application for the encryption key, which responds with a request to create the SPC. FPS verifies the content provider’s public key certificate and encrypts the SPC with the provider’s RSA public key. FPS also uses a challenge-response mechanism and an undisclosed \enquote{symmetric algorithm} \cite{FairPlayStreamingOverview2016a} to authenticate the key server. The player application creates a session with the key server and sends the encrypted SPC along with the content ID to the key server.

The key server decrypts the SPC using its RSA private key and creates a CKC. The CKC contains the IV, the encryption key and its expiration time, which is based on the player application’s time reference instead of the server’s time location. The key server encrypts the CKC with the session key contained in the SPC and sends it to the FPS on the device through the player application \cite{FairPlayStreamingOverview2016a}. 

The FPS creates a play context and uses the iOS kernel to decrypt the CKC using the session key. The content encryption key and the IV are extracted from the CKC, and content is decrypted on a per frame basis. The decrypted content is sent directly to the decoder by the FPS and HDCP is enforced to display the decoded content on device display \cite{FairPlayStreamingOverview2016a}.

\subsubsection{Server Playback Context (SPC)}
The SPC created by FPS contains various security elements, including:
\begin{itemize}
\item A random AES-128 session key for single SPC use; a new session key is required even if another SPC contains the same content ID.
\item An anti-replay seed to prevent replay of key server responses.
\item A secure version of the content ID for server verification.
\item Optional integrity verification to protect session key exchange.
\item Player application certificate and a unique and anonymous device identifier; these are used by the key server to restrict the number of simultaneous playbacks for a user account. 
\item Player application’s time reference, which is used by the key server to manage expiration of the encryption keys.
\item Versions of security mechanisms on the key server that are supported by the player application.
\end{itemize}

\subsubsection{Key Expiration}
FPS uses two modes of key expiration: \emph{video rental} and \emph{secure lease}, which can be used together or independently. FPS does not decrypt the content and declines playback if the encryption key has expired for a video rental. For a mid-playback expiration, the current playback is allowed to continue and the expiration is enforced for the next playback session.  The key server restricts the number of simultaneous playbacks by a single user account for a secure lease. The key server uses the anonymous device identifier to allocate a slot to the device; if the lease is not renewed by the player application before its expiration then the slot is deallocated and FPS stops content playback on the device.

\subsubsection{Offline Playback}
The key server allows offline playback by indicating the persistent nature of the encryption key in the CKC. As such, FPS creates a persistent play context that is linked to the iOS device, which is stored in the device's file system with the expiration time of the encryption key.

\subsubsection{Rollback Prevention}
The SPC includes the key server versions supported by the player application. FPS uses the latest compatible version of security mechanisms available on the platform to protect encryption keys. An undisclosed \enquote{protection mechanism} \cite{FairPlayStreamingOverview2016a} is used to prevent roll-back attacks to an older version of the security mechanisms.

\subsubsection{AirPlay Streaming}
AirPlay streaming from an iOS device to Apple TV uses a streamer context that supports secure streaming of encrypted content and CKC over AirPlay. According to Apple's documentation \cite{FairPlayStreamingOverview2016a}, FPS over AirPlay uses the same level of protection as FPS on iOS devices for the delivery of encryption keys and IVs.

\subsection{Security Architecture of Microsoft PlayReady} \label{playready}

\begin{figure}
\centering
 	\includegraphics[width=\linewidth]{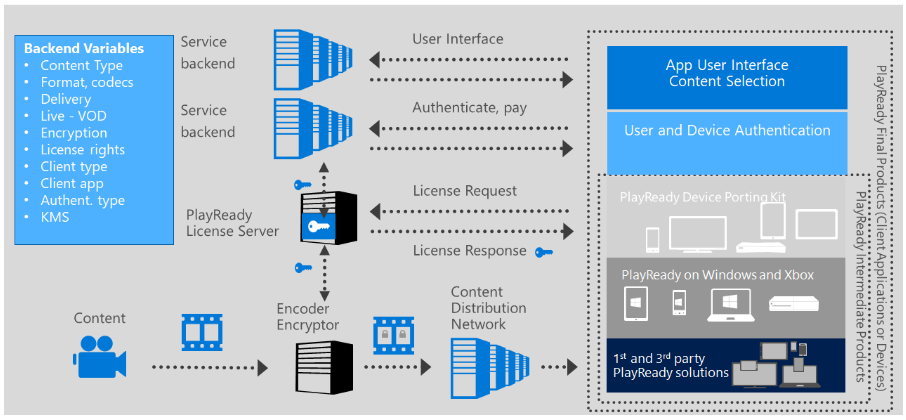}
 	\caption{Interactions between PlayReady client and servers \cite{sasouvanhPlayReadyOverviewSimple2018}.} 
 	\label{fig:playreadyendtoend} 
  \vspace{-0.2cm}
\end{figure}

PlayReady is widely deployed on Windows-based platforms, such as workstations and laptops, first developed in 2007 and currently integrated into Windows 10. PlayReady clients can be integrated into the OS or in the hardware. They can be implemented in media players on desktops, and as applications on mobile devices and other consumer devices, including smart TVs, set-top boxes and network receivers \cite{sasouvanhPlayReadyOverviewSimple2018}. PlayReady serves more as an end-to-end framework than the DRM solutions discussed hitherto, which provide more rigid requirements. Indeed, it does not provide any mechanisms for content encryption, user authentication, and key management; content providers are required to implement these features. However, it does employ a key seed, a random 30-byte value that content providers can use to derive encryption keys or they can use their own key management system. PlayReady supports MPEG-DASH, HLS and Microsoft Smooth Streaming protocols, and CENC Common Encryption using either AES-CBC or AES-CTR modes of operation \cite{vijayanagarWhatMicrosoftPlayReady2020}.  Content providers can use the PlayReady Server Software Development Kit (SDK) to build various servers that are compatible with PlayReady clients under a flexible model. These servers include content packaging servers, web servers, license servers, metering servers, and domain controllers.





\subsubsection{PlayReady Playback Security}  
Interactions between PlayReady-enabled clients and servers are illustrated in Figure \ref{fig:playreadyendtoend}. Here, encrypted content is distributed through a CDN for streaming to clients. The packaging server includes the KeyID for the content encryption key in the content header, according to the PlayReady header specification.

\begin{figure}
\centering
 	\includegraphics[width=\linewidth]{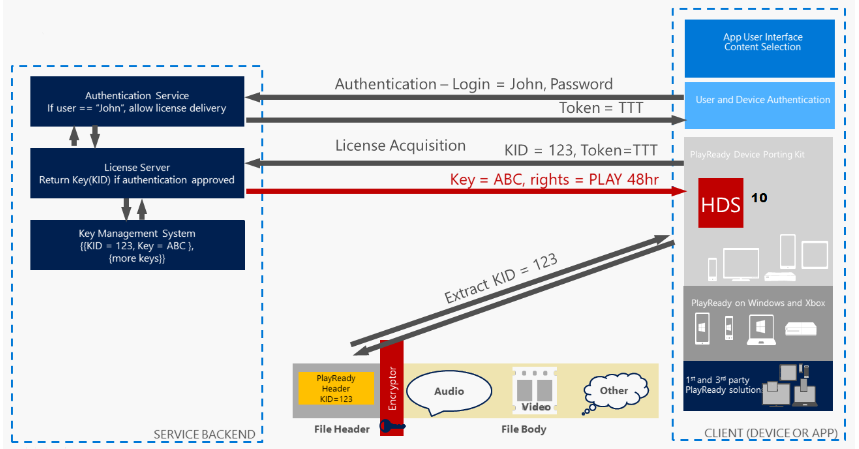}
 	\caption{Acquiring a PlayReady license \cite{sasouvanhPlayReadyOverviewSimple2018}.} 
 	\label{fig:playreadylicense} 
  \vspace{-0.2cm}
\end{figure}

The client streams the selected content from the CDN after browsing through the user interface. The client extracts the KeyID from the content header and obtains an authentication token from the authentication server. Figure \ref{fig:playreadylicense} details the process flow to obtain a license from the license server. The client sends a license request to the license server including the KeyID and the authentication token. The license server verifies the authentication token with the authentication service; retrieves the encryption key from the key management system; creates a license containing the encryption key and content usage rights and restrictions; encrypts the license using the client’s public key or the client domain public key; digitally signs the license with its own private signing key; and sends the signed license to the client. Multiple licenses and encryption keys can be sent in a single transaction and a license can also be granted for a domain instead of a single client \cite{sasouvanhPlayReadyOverviewSimple2018,sasouvanhPlayReadyOverviewLicense2018}.

The PlayReady client verifies the License server’s digital signature and certificate expiration, and then decrypts the license using its own private key \cite{sasouvanhPlayReadyOverviewLicense2018}. The client decrypts the content using the encryption keys from the license and it enforces the license policy. PlayReady client implementations are expected to follow PlayReady's compliance and robustness rules, which govern how clients decrypt the content and require the clients to correctly process and apply the restrictions in the license~\cite{sasouvanhPlayReadyOverviewSimple2018}.

\subsubsection{Offline Playback}
The license server can specify a particular license to be persistent or non-persistent. Persistent licenses are stored in non-volatile memory, such as the hard drive, and non-persistent licenses are stored in volatile memory so they are discarded when the current session ends \cite{sasouvanhPlayReadyOverviewSimple2018,sasouvanhPlayReadyOverviewLicense2018a}. The HDS (Hashed Data Store), shown in figure \ref{fig:playreadylicense}, is a client-side license store provided by PlayReady.

\subsubsection{PlayReady Security Levels}
The security level of a PlayReady client is included in the client certificate that is embedded in the client during manufacturing \cite{sasouvanhPlayReadyOverviewSecurity2018}. A copy of the client certificate in included in the license request and the license server can select suitable licenses and license properties depending on the client security level. Moreover, the license server sets a minimum required security level for each license; the security level of a client must exceed this threshold to be able to use the license. The PlayReady SL150 security level is the least secure and it is only meant to be used for development and testing of clients as the content and encryption keys are not protected. The SL2000 security level is for commercial client implementations in hardened devices and applications, which use software-based protection mechanisms. The SL3000 security level offers the highest security and it is designed for commercial implementations on hardened devices that use hardware-based protection mechanisms, utilising the TEE for core functionalities.

\subsection{Discussion}
\label{features}

In Table~\ref{tab:cfeatures}, we summarise the features provided by Widevine, FairPlay and PlayReady. We observe that none of these technologies provide all of the stated features based on publicly available information, placing an onus on content providers to implement any shortcomings. For instance, we note that no solution provides user authentication, while metering facilities are provided only by PlayReady. Conversely, all three DRM systems provide creation of license requests, license encryption, content decryption in the TEE of the device, and secure key storage for offline playback.  License request encryption and a secure display are provided by Widevine and FairPlay, but not PlayReady. Player application, license server and content packaging are provided by Widevine and PlayReady but not FairPlay; PlayReady does not provide the actual mechanisms, it has the client compliance and robustness rules for player applications, and a server SDK to develop packaging and license servers.  Content encryption is only provided by Widevine using the Shaka packager. Only FairPlay authenticates the license server and anonymously limits simultaneous playback sessions. Features unique to PlayReady include a key seed for key generation, and server SDK to develop web servers, domain-based license servers and metering servers.

\begin{table}
\begin{threeparttable}
\caption{Feature fulfilment of Widevine, FairPlay, and PlayReady.}
\centering

\begin{tabular}{ | p{0.12\textwidth} | p{0.1\textwidth} | p{0.1\textwidth} | p{0.08\textwidth} | }

 \hline
  \textbf{Feature} & \textbf{Widevine} & \textbf{FairPlay} & \textbf{PlayReady} \\ 
  \hline
 \textbf{Key generation} & --- & --- & Key seed \\
  \hline
  \textbf{Packaging} & Shaka packager & --- & PlayReady server SDK \\
 \hline
  \textbf{Encryption} & Shaka packager & --- & --- \\
  \hline
  \textbf{User authentication} & --- & --- & --- \\
  \hline
  \textbf{License server authentication} & --- & RSA-based protocol & --- \\
  \hline
  \textbf{Create license request} & Widevine CDM & FPS framework & PlayReady client \\
  \hline
  \textbf{License request encryption} & Encrypted & Encrypted with key server's RSA public key & --- \\
  \hline
  \textbf{License server} & Widevine license server & --- & Server SDK \\
  \hline
  \textbf{License encryption} & Encrypted & Session key and anti-replay & Client/domain public key, digitally signed \\
  \hline
  \textbf{Content management} & --- & --- & --- \\
  \hline
  \textbf{Content storage} & --- & --- & PlayReady server SDK \\
  \hline
  \textbf{Player application} & Widevine media player & --- & Client compliance rules \\
  \hline
  \textbf{TEE component} & OEMCrypto module & FPS framework & PlayReady client \\
  \hline
  \textbf{Display} & EME/MSE & HDCP enforced & --- \\
  \hline
  \textbf{Offline playback} & Widevine keybox & iOS file system & HDS (Hashed Data Store) \\
  \hline
  \textbf{Limit playback sessions} & --- & Anonymous device ID & --- \\
  \hline
  \textbf{Domain license} & --- & --- & Server SDK \\
  \hline
  \textbf{Metering} & --- & --- & Server SDK \\
  \hline
\end{tabular}
\label{tab:cfeatures}
\begin{tablenotes}
    \item ---: Specific measures not known from publicly available documentation.
\end{tablenotes}
\end{threeparttable}
\vspace{-0.3cm}

\end{table}

\section{DRM evaluation methodology and criteria} \label{evaluation}
In this section, we distil the features and security properties of cloud-based DRM systems for mobile devices.\footnote{We note that one possible approach is to employ a threat modelling methodology, such as STRIDE \cite{STRIDEThreatModel2009a} or NIST SP 800-154 \cite{souppayaDraftNISTSpecial2016}. These approaches identify vulnerabilities and potential threats using data flow diagrams and attack trees; threats are then ranked and appropriate mitigations are determined.  Generic threat modelling methodologies, however, do not specifically cover the features necessary for constructing secure DRM systems, thereby warranting a domain-specific evaluation.}  

\subsection{DRM Security Features} \label{taxonomy}

Based on the previous sections, we propose a general feature set for classifying OTT DRM systems for cloud-based content delivery to mobile devices. The set assumes an end-to-end workflow of security processes and types of vulnerabilities specific to cloud-based mobile DRM systems, comprising 21 key security aspects listed under the following categories (also given in Table \ref{table:taxonomy}). The following sections describe the various security aspects, numbered \ref{tax(1)}--\ref{tax(21)}.

\begin{table}

\centering
\caption{Security properties of cloud-based mobile DRM systems.}
\makebox[\textwidth][l]{
\begin{tabular}{ | p{0.10\textwidth} | p{0.12\textwidth} | p{0.1\textwidth} | p{0.06\textwidth} | }
 \hline
  \textbf{Category} & \multicolumn{2}{c|}{\textbf{Subcategories}} & \textbf{Property} \\
 \hline
  \multirow{6}{0.2\textwidth}{\textbf{Key\\management}} & \multicolumn{2}{l|}{\emph{Key generation}} & \ref{tax(1)} \\
  \cline{2-4}
  & \multirow{2}{0.2\textwidth}{\emph{Key in transit}} & \emph{Pre-encryption} & \ref{tax(2)} \\
  \cline{3-4}
  & & \emph{Post-encryption} & \ref{tax(3)} \\
  \cline{2-4}
  & \multicolumn{2}{l|}{\emph{Key storage}} & \ref{tax(4)} \ref{tax(5)} \\
  \cline{2-4}
  &  \multicolumn{2}{l|}{\emph{Key disposal}}  & \ref{tax(6)} \\
 \hline

  \multirow{4}{0.2\textwidth}{\textbf{Data\\encryption}} & \multirow{2}{0.2\textwidth}{Content encryption} & \multicolumn{1}{l|}{\emph{Standardisation}} & \ref{tax(7)} \\
  \cline{3-4}
  & & \multicolumn{1}{l|}{\emph{Re-keying}} & \ref{tax(8)} \\
  \cline{2-4}
  & \multirow{2}{0.2\textwidth}{\emph{Key encryption}} & \multicolumn{1}{l|}{\emph{Session security}} & \ref{tax(9)} \\
  \cline{3-4}
  & & \multicolumn{1}{l|}{\emph{Public-key crypto.}} & \ref{tax(10)} \\
  \hline
  %
  %
  %
  \multirow{5}{0.2\textwidth}{\textbf{Content\\management}} & \multirow{2}{0.2\textwidth}{\emph{Content storage}} & \emph{Pre-encryption} & \ref{tax(11)} \\
  \cline{3-4}
  & & \emph{Post-encryption} & \ref{tax(12)} \\
  \cline{2-4}
  & \multirow{3}{0.2\textwidth}{\emph{Content in\\transit}} & \emph{Pre-encryption} & \ref{tax(13)} \\
  \cline{3-4}
  & & \emph{Post-encryption} & \ref{tax(14)} \\
  \cline{3-4}
  & & \emph{Post-decryption} & \ref{tax(15)} \\
  \cline{2-4}
  & \multicolumn{2}{l|}{\emph{Manifest security}} & \ref{tax(16)} \\
 \hline

  \multirow{2}{0.2\textwidth}{\textbf{Authentication}} & \multicolumn{2}{l|}{\emph{User authentication}} & \ref{tax(17)}\\\cline{2-4} 
  & \multicolumn{2}{l|}{\emph{Server authentication}} & \ref{tax(18)}  \\
  \hline

  \textbf{DDoS} & \multicolumn{2}{l|}{\emph{DDoS mitigations}} & \ref{tax(19)} \\\hline

  \textbf{TEE} & \multicolumn{2}{l|}{\emph{Secure TEE usage}} &  \ref{tax(20)} \\ 
 \hline

  \textbf{Quantum security} & \multicolumn{2}{l|}{\emph{Post-quantum mitigations}} & \ref{tax(21)} \\\hline
\end{tabular}
}

\label{table:taxonomy}
\end{table}

\subsubsection{Key management} \label{keymanagement}
In mobile DRM systems, cryptographic keys and IVs must be generated securely, and keys must also be protected in transit (to the packaging and license servers) and storage (in a key database). The packaging server uses the keys to encrypt content, while the license server embeds the keys in an encrypted license which is transported to the TEE, where it is used to support content decryption.

\begin{enumerate}[start=1,leftmargin=1cm,label={[SP\arabic*]}]
\item \textbf{Key generation:} \label{tax(1)} a secure random number generator (RNG), or a key seed, is required to provide unpredictable values for generating keys and IVs with sufficient entropy.

\item \textbf{Pre-license key transit:} \label{tax(2)} secure channels are required to transfer keys to the packaging and license servers.

\item \textbf{License transit:} \label{tax(3)} keys are securely embedded in an encrypted and authenticated license.

\item \textbf{Key database:} \label{tax(4)} keys are securely held in a key database before encryption by the license server.

\item \textbf{On-device key storage:} \label{tax(5)} keys should be stored in TEE-reserved memory regions on user devices.

\item \textbf{Key disposal:} \label{tax(6)} keys should be disposed securely to avoid content decryption once the license has expired.
\end{enumerate}

\subsubsection{Encryption} \label{taxonomyencrypt}
DRM systems commonly employ symmetric keys for content encryption, which are protected using key encryption keys. Public key cryptography is used for license encryption, containing content encryption keys, and for exchanging (symmetric) session keys. 

\begin{enumerate} [resume,leftmargin=1cm,label={[SP\arabic*]}]
\item \textbf{Suitable encryption algorithms:} \label{tax(7)} Standard symmetric algorithms are required for secure and efficient content encryption, e.g.\ 128-bit AES. 

\item \textbf{Content re-keying:} \label{tax(8)} refers to encrypting different parts of the content with different keys to limit the impact of a key compromise, and to allow different license rules to be applied to each part.

\item \textbf{Session key management:} \label{tax(9)} session keys must be generated, exchanged, and stored securely.

\item \textbf{Public key management:} \label{tax(10)} public-key pairs must be generated and managed securely using appropriate standards and modules, e.g.\ NIST FIPS 140-3.
\end{enumerate}

\subsubsection{Content management} \label{contentmanage}
Content and the manifest files require secure storage and transit. Content is transferred from the content database to the packaging server to be encrypted before it is stored in a content server and cached by the CDN. The encrypted content is streamed to user devices, decrypted in the TEE, and sent to the device display for viewing.

\begin{enumerate} [resume,leftmargin=1.2cm,label={[SP\arabic*]}]
\item \textbf{Pre-packaged content storage:} \label{tax(11)} Content requires secure storage in the content database to prevent attackers from obtaining plaintext content before DRM protection is applied.

\item \textbf{Packaged content storage:} \label{tax(12)}
Packaging servers generally apply encryption during content packaging.

\item \textbf{Pre-packaged content transit:} \label{tax(13)} content requires secure transfer from the content database to the packaging server where it is encrypted.

\item \textbf{Packaged content transit:} \label{tax(14)} Packaged content is encrypted for transit.

\item \textbf{Secure video path:} \label{tax(15)} decrypted content is displayed directly from the TEE to the display device over a trusted path, e.g.\ to thwart screen scraping attacks.

\item \textbf{Secure manifest:} \label{tax(16)} manifest files, containing URLs for different content segments, must be transmitted securely. A MITM attack, where the license server or CDN is impersonated, could potentially replace these URLs with malicious ones. 
\end{enumerate}

\subsubsection{Authentication} \label{accesscontrol} 
Mutual authentication between users and the servers is required to ensure only authorised users can access the content, and to prevent false server attacks, where an attacker-controlled server impersonates a legitimate server to launch attacks on user devices.

\begin{enumerate} [resume,leftmargin=1.2cm,label={[SP\arabic*]}]
\item \textbf{User authentication:} \label{tax(17)} users should be authenticated by the CDN and the license server.

\item \textbf{Server authentication:} \label{tax(18)} CDN and license servers should be authenticated by user devices.
\end{enumerate}

\subsubsection{DDoS} \label{ddos}
Compromise the availability of DRM-protected content by consuming server and network resources.

\begin{enumerate} [resume,leftmargin=1.2cm,label={[SP\arabic*]}]
\item \textbf{DDoS vulnerabilities:} \label{tax(19)} CDN and license servers could potentially be vulnerable to DDoS attacks.
\end{enumerate}

\subsubsection{TEEs} \label{tee}
DRM protection relies on the security provided by TEEs on user devices. Vulnerabilities in TEEs could be leveraged to compromise DRM-protected content.

\begin{enumerate} [resume,leftmargin=1.2cm,label={[SP\arabic*]}]
\item \textbf{TEE attack resistance:} \label{tax(20)}
TEEs share hardware resources, such as the processor and memory, with the native OS on user devices. A TEE must be able to thwart relevant attacks, such as side-channel and fault injection attacks. 
\end{enumerate}

\subsubsection{Quantum security} \label{quantum}
Suitable post-quantum algorithms must be used to maintain the security of DRM systems against quantum adversaries, which can undermine public- and symmetric-key cryptographic algorithms.

\begin{enumerate} [resume,leftmargin=1.2cm,label={[SP\arabic*]}]
\item \textbf{Post-quantum security:} \label{tax(21)} the outcomes of NIST's Post-quantum Cryptography (PQC) standardisation process must be implemented. 
\end{enumerate}

\section{Security Analysis of DRM Technologies} \label{analysis}


In this section, we consider the highest security levels of Widevine, FairPlay and PlayReady, e.g.\ Widevine L1 and PlayReady SL3000 levels, which utilise hardware-assisted mechanisms for content protection.  Widevine, FairPlay and PlayReady provide a subset of the security features (\S\ref{features}) required for a complete DRM implementation; content providers must implement the remaining features securely (\S\ref{secureimplement}) for the DRM protection to be effective. We analyse each technology in turn below, and compare them in Table~\ref{table:comparison}. 



\textbf{Widevine}. We observe that Widevine poses security risks due to the way in which manifest files could be used as an attack vector. Content encryption keys in Widevine are at a low risk during license transit, as the license is encrypted and sent over HTTPS. Risks to the encryption algorithm and packaged content, both in storage and in transit, are low as the Shaka packager uses CENC encryption to encrypt the content. The risk level of TEE-related vulnerabilities is low due to the specialist equipment and skills required for these attacks. The risk from post-quantum attacks is present, albeit low given the absence of large-scale quantum computers. New post-quantum public key algorithms will be required to replace the current ones, and the current symmetric key lengths will not offer sufficient protection from quantum attacks.


\textbf{FairPlay}. Like Widevine, FairPlay poses a high risk to user devices due to insecure manifest files. FairPlay requires the license to be encrypted with a unique AES-128 session key; generated by the FairPlay client; protected by the license server’s public key; and also protected from replay attacks by an anti-replay seed. Therefore, the security risk to the license during its transit is low. Content providers are required to encrypt the packaged content using AES-128 CBC mode and a unique IV for each encryption. Unlike other TEEs, Apple's Secure Enclave Processor (SEP)~\cite{shepherd2021physical} does not share hardware resources with host OS on the device and protects against some side-channel attacks, including attacks based on cache-timing differences. It is unclear whether the SEP is used by FairPlay, although we recommend that it should be used for additional hardware-assisted protection. We note that FairPlay is also vulnerable to quantum adversaries.


\textbf{PlayReady}. The PlayReady manifest files also pose a high risk to user devices. The license is encrypted with the PlayReady client public key or the client domain public key, and it is digitally signed by the license server; hence, the license in transit has a low risk level. The PlayReady client verifies the digital signatures and decrypts the license using its private key; the level of risk to the public-private key pairs is low. PlayReady supports CENC Common Encryption with AES-CBC and AES-CTR for content encryption, resulting in a low risk to packaged content in storage and in transit. There is a risk to PlayReady from TEE and quantum adversaries.

\begin{table}
\centering
\caption{Security Comparison of Widevine, FairPlay, and PlayReady.}
\begin{threeparttable}
    
\begin{tabular}{  | p{0.23\textwidth} | p{0.055\textwidth} | p{0.05\textwidth} | p{0.065\textwidth} | }
 \hline
  \multirow{2}{0.36\textwidth}{\textbf{Security Property}}& \multicolumn{3}{c|}{\textbf{Technology}} \\
 \cline{2-4}
   & \textbf{Widevine} & \textbf{FairPlay} & \textbf{PlayReady} \\
 \hline
  \ref{tax(1)} Key generation & \amark & \amark & \amark \\
 \hline
  \ref{tax(2)} Pre-license key transit & \amark & \amark & \amark \\
 \hline
  \ref{tax(3)} License transit & \cmark,\lnmark & \cmark,\lnmark & \cmark,\lnmark \\
 \hline
  \ref{tax(4)} Key database & \xmark& \xmark& \xmark\\
 \hline
  \ref{tax(5)} Key storage on device & \amark & \amark & \amark \\
 \hline
  \ref{tax(6)} Key disposal & \xmark& \amark & \amark \\
 \hline
  \ref{tax(7)} Suitable encryption algorithms & \cmark,\lnmark & \cmark,\amark & \cmark\amark \\
 \hline
  \ref{tax(8)} Content re-keying & \xmark& \xmark& \xmark\\
 \hline
  \ref{tax(9)} Session key management & \amark & \cmark,\lnmark & \xmark\\
 \hline
  \ref{tax(10)} Public key management & \amark & \xmark& \cmark,\lnmark \\
 \hline
  \ref{tax(11)} Pre-packaged content storage & \xmark& \xmark& \xmark\\
 \hline
  \ref{tax(12)} Packaged content storage & \cmark,\lnmark & \cmark,\lnmark & \cmark,\lnmark \\
 \hline
  \ref{tax(13)} Pre-packaged content transit & \xmark& \xmark& \xmark\\
 \hline
  \ref{tax(14)} Packaged content transit & \cmark,\lnmark & \cmark,\lnmark & \cmark,\lnmark \\
 \hline
  \ref{tax(15)} Secure video path & \smark & \amark & \amark \\
 \hline
  \ref{tax(16)} Secure manifest & \smark,\upmark & \smark,\upmark & \smark,\upmark \\
 \hline
  \ref{tax(17)} User authentication & \xmark& \xmark& \xmark\\
 \hline
  \ref{tax(18)} Server authentication & \xmark& \amark & \xmark\\
 \hline
  \ref{tax(19)} DDoS security & \amark & \xmark& \xmark\\
 \hline
  \ref{tax(20)} TEE security & \cmark,\lnmark & \amark & \cmark,\lnmark \\
 \hline
  \ref{tax(21)} Quantum security & \smark,\upmark & \smark,\upmark & \smark,\upmark \\
 \hline
\end{tabular}
\begin{tablenotes}
    \item \cmark: Provided natively. \xmark: Not provided. \smark: Somewhat provided. \amark: Insufficient information available. \lnmark: Contains security weaknesses of low concern. \upmark: Weaknesses of high concern.
\end{tablenotes}
\vspace{-0.4cm}
\end{threeparttable}

\label{table:comparison}
\end{table} 

\section{Mitigations} \label{mitigations}
Following the previous analysis, this section proposes mitigations for the identified issues.

\subsection{Malicious Manifest Attacks} \label{mitigatemanifest}
There is a high risk of potential attacks on user devices from malicious manifest files used in Widevine, FairPlay and PlayReady. The manifest file is sent as plaintext, and it contains DRM client initialization data, KeyIDs, and download locations of content segments for a content stream. Manifest files are sent as plaintext to allow DRM clients to stream relevant content segments defined in the manifest, and to initialize a license request using the KeyIDs contained in the manifest. Moreover, an encrypted manifest can be replaced with a malicious one by an attacker without requiring decryption. Data origin authentication of the manifest, as opposed to confidentiality, is required to mitigate these attacks. Content providers do not share long-term secrets with DRM clients to allow symmetric cryptography, such as MACs (Message Authentication Codes), to be used. Instead, content providers can digitally sign the manifest using a private signing key to provide non-repudiation, a stronger notion of data origin authentication, which may be accompanied by mutual authentication and TEE-based attestation (e.g.\ see \cite{shepherd2019remote});  DRM clients can use the content provider’s public verification key to verify the digital signatures before processing the manifest. 

\subsection{Secure Configuration} \label{secureimplement}
The overall security of a DRM system depends on its complete configuration; Widevine, FairPlay and PlayReady provide a subset of the features required for DRM protection (\S \ref{analysis}), leaving the remaining features to be implemented by content providers. An insecure implementation of these features could potentially introduce vulnerabilities in a DRM system. DRM systems use encryption to protect content and the content encryption keys. A secure implementation of standards-based encryption with correct key management should be used to ensure confidentiality of the content and the keys. 


\subsection{Authenticated Encryption}
Similar to the malicious-manifest attacks (\S \ref{mitigatemanifest}), the encrypted content segments and licenses can be replaced by malicious payloads or malware in a MITM attack. Data origin authentication of the content segments and licenses is required to prevent these attacks. Confidentiality of content segments and licenses is already provided by the packaging and license servers, which can be combined with data origin authentication protocols. There are several modes of operation for authenticated encryption that securely combine data origin authentication and confidentiality into a single primitive, such as CCM mode, EAX mode, OCB mode and GCM mode. Encryption of content segments by the packaging server, and licenses by the license server, should be replaced with authenticated encryption to protect the content and licenses in transit, and to protect user devices from MITM attacks described above.

\subsection{Trusted Execution Environments} \label{mitigatetee}
DRM systems use TEEs in user devices to securely decrypt and store content, and to manage cryptographic keys. In recent years, TEEs have been subjected to a myriad of side-channel attacks, particularly micro-architectural ones, as well as physical fault injection attacks. Some side-channel attacks, such as attacks based on cache-timing differences, are caused by hardware (processor and memory) sharing between the TEE and the host OS on user devices; these can be mitigated by using an isolated processor that is not shared with the main OS. Using an isolated processor for TEE could prevent cache-based and transient execution attacks that exploit cache memory sharing between the \enquote*{normal} and the \enquote*{secure} worlds to leak sensitive TEE data~\cite{xiong2021survey}. However, it does not prevent fault injection and other physical attacks, like those based on electromagnetic and power analysis. For this, tamper-resistant packaging is a widely suggested countermeasure~\cite{shepherd2021physical}.

\subsection{Post-quantum Mitigations}

Quantum computers are known to fatally undermine traditional public key algorithms, e.g.\ RSA and ECDSA, which has prompted the NIST PQC standardisation competition \cite{PostQuantumCryptographyStandardization2023}. To make DRM systems resistant to post-quantum attacks, TEEs should be adapted to the new PQC standards. The various security services provided by TEEs, such as authenticated boot, remote attestation and key management, lack widespread resistance to quantum adversaries, as do the supporting protocols for license distribution in existing DRM systems. As such, we strongly urge developers to expedite the transition to post-quantum algorithms for license distribution and authentication in particular.

\subsection{DRM Standardisation}
The fragmentation in the current DRM landscape  and lack of standardisation is a source of potential vulnerabilities in DRM systems. The DRM systems analysed in this work use different formats, protocols and APIs. The complexity of the ecosystem, with different implementations using numerous components, may inadvertently lead to insecure configurations. Additionally, various standardised formats and APIs for DRM systems have been developed, including CENC \cite{ISOIEC2300172016}, CMAF (Common Media Application Format) \cite{ISOIEC23000192020}, CPIX (Content Protection Information Exchange) \cite{DASHIFImplementationGuidelines2018} and EME APIs \cite{dorwinEncryptedMediaExtensions2017}. This paper has shown that some of these standards are not used by any of the DRM systems, such as the CPIX, a key exchange data format for confidential and authenticated key delivery in DRM systems. Further DRM standardisation could potentially improve interoperability, reduce fragmentation and complexity, and minimise vulnerabilities.

\section{Future Directions} \label{future}

The centralised servers in cloud-based DRM can be a single point of failure and vulnerable to attacks. In a centralised DRM model, content owners do not have direct access to consumers and must sell their content to large content providers and media platforms. Furthermore, the copyright and license transaction information in current DRM systems lacks transparency, and authentication is inefficient, requiring multiple interactions. To overcome these challenges, Zhang and Zhao \cite{zhangDesignDigitalRights2018a} have proposed a decentralised DRM model with collective maintenance using a blockchain-based license structure. Licenses are issued automatically using smart contracts, offering reliability of license transactions, and the license transaction information is recorded on the blockchain to make it transparent and secure. The authorisation of licenses is scalable and does not need interactive communication between content owners and consumers. Content owners can set flexible pricing rules for content usage and deliver DRM-protected content directly to the consumers. 

Rana and Mishra~\cite{ranaSecureUbiquitousAuthenticated2020a} have proposed an authenticated key distribution framework for IoT enabled DRM systems, allowing authorised users to consume content through smart devices. The framework uses biometric data and three-factor authentication to verify users; allows secure and anonymous communication; but does not preserve the secrecy of consumer preferences. Data centres consume a vast amount of power to meet consumer demand for low-latency access to content. Rana et al.~\cite{ranaComputationalEfficientAuthenticated2020} have discussed the security of authenticated DRM key distribution frameworks, which support efficient communication, computing and mutual authentication to conserve energy consumption. Lastly, we draw attention to Li et al.~\cite{liCloudBasedVideoStreaming2020}, who explored using federated cloud and edge computing for cost-efficient streaming, and developing load distribution mechanisms to reduce latency.

\section{Conclusion} \label{conclusion}

In this paper, we presented the first security evaluation of three major DRM systems under the OTT content delivery model with respect to mobile devices.  The security analysis identified attack vectors in all aspects of key management, content management, encryption and access control in each of the DRM systems, drawing from parallel advancements in the state of the art (e.g.\ TEE micro-architectural vulnerabilities). We showed how all three DRM systems are vulnerable to post-quantum attacks, and issues pertaining to manifest security. Moreover, TEEs are vulnerable to a range of side-channel and physical fault injection attacks that could compromise DRM-protected content. However, it is worth being cognisant of the high levels of expertise and specialist equipment needed to mount such attacks successfully. Another source of potential vulnerabilities is the complexity and fragmentation in the DRM landscape, which could be remedied by appropriate standardisation efforts to reduce the fragmentation and complexity of the ecosystem. Some of the current trends and future directions of research relevant to cloud-based DRM include a decentralised DRM model based on blockchain technology; IoT-enabled DRM systems; improving the power and computational efficiency of DRM data centres; and using federated cloud and edge computing with load distribution mechanisms for cost-efficient and low latency streaming.


\bibliographystyle{IEEEtran}
\bibliography{IEEEabrv, refs}

\end{document}